\documentstyle[twocolumn,aps,prl,epsf]{revtex} 
\begin{document}

\twocolumn[\hsize\textwidth\columnwidth\hsize\csname      
@twocolumnfalse\endcsname

\draft
\title{On the three-body continuum spectrum of $^6$He}
\author{Attila Cs\'ot\'o}
\address{Department of Atomic Physics, E\"otv\"os University, 
Puskin utca 5-7, H--1088 Budapest, Hungary}
\date{July 6, 1998}

\maketitle

\begin{abstract}
\noindent
In recent publications Cobis, Fedorov, and Jensen claim the
existence of several previously unknown low-lying narrow
resonances in $^6$He. I show that the distribution of the
S-matrix poles corresponding to these states is unphysical. 
This casts doubt on the results of those works concerning
resonances. 
\end{abstract}
]

\narrowtext

Recently a series of papers has been published on three-body
continuum calculations for the neutron-halo nuclei $^6$He and
$^{11}$Li \cite{Cobis}. The authors report on numerous
previously unknown states they find in these nuclei at low
energies. Most importantly they claim to have found the much
debated soft dipole resonances in $^6$He and $^{11}$Li. I argue
that the presentation of Ref.\ \cite{Cobis} is rather
misleading, because important information about the
calculations, that would have cast doubt on the results, 
were not mentioned. As I show, the results regarding the
existence of low-energy narrow states in $^6$He are highly
questionable.

Three-body resonances in an A=6 nucleus, $^6$Li, were first
studied in Ref.\ \cite{Eska} in an $\alpha+p+n$ model. A
systematic search for such states in $^6$He, $^6$Li, and $^6$Be
was first performed in Ref.\ \cite{soft} with the aim to confirm
or refute the existence of the soft dipole ($1^-$) resonance in
$^6$He. Only the known states were found in the three nuclei,
and no evidence for the existence of the soft dipole resonance
in $^6$He surfaced. As was stated in Ref.\ \cite{soft}, the
method was numerically not stable enough for broad states
($\Gamma\gg E$), so the possible existence of such states cannot
be ruled out based on that work.

The $^6$He nucleus was studied also in an $\alpha+n+n$ model
with structureless $\alpha$ in Refs.\ \cite{RNBT,Aoyama} using
different interactions and different methods. They both find
several previously unknown states. While there is a good
agreement between Refs.\ \cite{RNBT,Aoyama} in the ordering and
spin-parities of the new states, the Ref.\ \cite{RNBT}
resonances are relatively narrow, whereas all new states in
Ref.\ \cite{Aoyama} are very broad. The relatively narrow states
found in Ref.\ \cite{RNBT} should have been seen in Ref.\
\cite{soft}, as the method used there was adequate for them.
However, neither the original work \cite{soft} nor new
calculations \cite{unpub} show such states. On the other hand,
the Ref.\ \cite{soft} model cannot rule out the broad states of
Ref.\ \cite{Aoyama}. We note that it is quite possible that
despite the big differences in the widths, the Ref.\ \cite{RNBT}
and Ref.\ \cite{Aoyama} states correspond to each other, and the
differences come mainly from the different Hamiltonians. Test
calculations using the same Hamiltonian would be desirable.

Finally, the latest experiments do not seem to support the
existence of any new narrow states in $^6$He \cite{Janecke}.

In contrast to all previous works which suggested only a few (if
any) new resonances in $^6$He, Ref.\ \cite{Cobis} predicts 
several rather
narrow states in each $J^\pi$ channel. The authors of Ref.\
\cite{Cobis} avoid the use of the words ``state'' or
``resonance'' in connection with the S-matrix poles they find.
However, one must realize that according to mathematical
theorems, for well-behaved potentials all poles of an S matrix
in the meromorphic region of the potential are physical, and 
correspond to resonances \cite{Newton}. Thus, if the mathematical 
conditions for the potentials are satisfied, the analytic 
continuation of the S matrix is done properly, and the whole 
procedure is numerically stable, then all the poles in Ref.\ 
\cite{Cobis} should correspond to real resonances of $^6$He.

In Fig.\ 1 the complex-energy positions of the first four
S-matrix poles found \cite{Cobisthesis} by the authors of Ref.\
\cite{Cobis} in the $J^\pi=1^-$ Hamiltonian are shown. Although
all these poles (and possibly more) were known to the authors,
they elected to show only the first two in each partial waves in
Ref.\ \cite{Cobis}. The distribution of the poles in Fig.\ 1 is
{\em clearly unphysical}, and thus some or all of them must
be artifacts. The origin of these spurious states can be threefold: i)
the ``effective potentials'' appearing in the method of
\cite{Cobis} do not satisfy the necessary mathematical
conditions; ii) the analytic continuation is not done with
sufficient care in \cite{Cobis}; iii) the whole method of Ref.\
\cite{Cobis} is questionable. Personally I think that point ii) 
is probably the (main) source of the problem. The three-body
problem has a rather complicated analytic structure at complex
energies especially if there are resonances in the two-body
subsystems, like in $\alpha+n+n$. The analytic structure of the
Riemann energy surface was discussed in detail, e.g., in Ref.\
\cite{Afnan}. One can see in Ref.\ \cite{Afnan} that, for
example, the resonant poles of the two-body subsystems appear as
complex-energy thresholds with two-body branch cuts in the
complex plane in the three-body problem (for a numerical
illustration, see also \cite{3br}). It means that for
$\alpha+n+n$ there are branch cuts starting at the $\alpha+n$
pole energies ($0.77-i0.32$) MeV and ($1.97-i2.61$) MeV of Ref.\
\cite{Cobis,Cobisthesis}, respectively for $J^\pi=3/2^-$ and
$1/2^-$. It seems rather plausible that the two highest-energy
points in Fig.\ 1 are part of the $3/2^-$ cut. If this is so,
then further ``poles'' should be found lying on this line closer 
to the imaginary $E$ axis. 

In order to be able to understand the nature of the first two
poles, one should know more about the details of the analytic
continuation used in Ref.\ \cite{Cobis}. We mention just one
example that the authors used rather unorthodox conventions:
they defined the three-body branch cut along the {\em negative}
real energy axis, thus mapping the left and right half
$k$-planes onto a Riemann surface instead of the top and 
bottom half-planes. This may seem just a matter of choice, 
but it might violate some fundamental symmetries as well. 

In Ref.\ \cite{Cobis} it is shown that the asymptotic part of
the wave function for the $1^-$ state starts at very large $r$.
This can imply that other methods might not be able to handle
this behavior correctly, and as a consequence might miss the
$1^-$ state. This is a valid argument, so we checked it in the
complex scaling method used in Ref.\ \cite{soft}. The $1^-$
state did not appear even if the range of the basis extended
beyond 100 fm. At the same time the position of the $2^+$ state
remained remarkably stable despite the fact that such a basis is
numerically unfavorable.

An implicit argument in Ref.\ \cite{Cobis,Cobisthesis}, to support the
existence of the low-lying states in $^6$He, is the attractive
nature of both the $n+n$ and $\alpha+n$ forces (in other words,
the large scattering lengths) in the crucial partial waves.
Although this may seem to be a logical argument, there is at 
least one well-known counter-example to it: the nonexistence of 
the $1/2^+$ state of the three-neutron system \cite{Glockle,3n}. 
The most important configuration of such a state would be an 
$L=0$ relative motion between a $^1S_0$ dineutron and the third
neutron. This would contain the attractive $^1S_0$ $N-N$
interaction in all two-body subsystems. Yet there is no
evidence, either theoretical or experimental, for the existence
of such a state. I think that a good test of the methods of
Ref.\ \cite{Cobis} would be the $3n$ system.

Finally, I would like to emphasize that a future measurement of
some real-energy observables, like the dipole strength function,
cannot be used as a proof for the existence of the states in
Ref.\ \cite{Cobis}, even if the data happened to agree with the 
theoretical prediction. As it was shown in Ref.\ \cite{soft,3br}, 
some structures of the three-body continuum can come not only 
from three-body resonances but, for example, from the two-body
substructure. The sequential breakup of $^6$He might produce a
strength function similar to those in Ref.\ \cite{Cobis} without
any $1^-$ three-body resonance.

In conclusion, I have pointed out that the distribution of the
complex S-matrix poles in Ref.\ \cite{Cobis,Cobisthesis} is
clearly unphysical. It seems that most of the complex-energy
results of Ref.\ \cite{Cobis} might be artifacts caused by the
unsatisfactory handling of the analytic continuation. I think it
would be very beneficial if the authors of Ref.\ \cite{Cobis} tried 
to compare their results to other published ones, e.g. to Ref.\ 
\cite{Aoyama}, using the same potentials (as it should have been 
done at least for tests in the original work).

This work was supported by OTKA Grant F019701 and by the Bolyai
Fellowship of the Hungarian Academy of Sciences. I thank I.~R.
Afnan, S. Aoyama, C. Chandler, A. Cobis, D.~V. Fedorov, A.~S. 
Jensen, K.  Kat\=o, and M.~V. Zhukov for useful discussions.

\begin{figure}
\epsfxsize 8cm \epsfbox{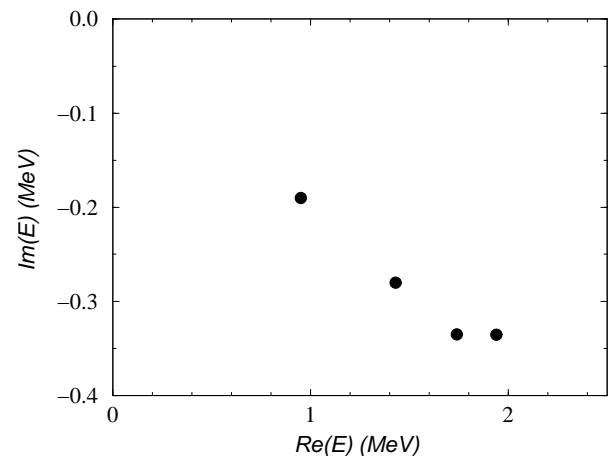}
\caption{Positions of the the first four poles of the 
$J^\pi=1^-$ S matrix of Ref.\ \protect\cite{Cobis,Cobisthesis} 
on the complex-energy plane.}
\end{figure}


\begin{references}
\bibitem{Cobis} A. Cobis, D.~V. Fedorov, and A.~S. Jensen, Phys.
Rev. Lett. {\bf 79}, 2411 (1997); Phys. Lett. {\bf B424}, 1
(1998); Nucl. Phys. {\bf A631}, c793 (1998); nucl-th/9804057.
\bibitem{Eska} A. Eskandarian and I.~R. Afnan, Phys. Rev. C {\bf
46}, 2344 (1992).
\bibitem{soft} A. Cs\'ot\'o, Phys. Rev. C {\bf 49}, 3035 (1994).
\bibitem{RNBT} B.~V. Danilin, T. Rogde, S.~N. Ershov, H. Heiberg-Andersen, 
J.~S. Vaagen, I.~J. Thompson, and M.~V. Zhukov, Phys. Rev. C {\bf 55},
R577 (1997); S.~N. Ershov, T. Rogde, B.~V. Danilin, J.~S. Vaagen, 
I.~J. Thompson, and F.~A. Gareev, Phys. Rev. C {\bf 56}, 1483
(1997); B.~V. Danilin, I.~J. Thompson, J.~S. Vaagen, and M.~V.
Zhukov, Nucl. Phys. {\bf A632}, 383 (1998).
\bibitem{Aoyama} S. Aoyama, S. Mukai, K. Kato, and K. Ikeda,
Prog. Theor. Phys., {\bf 93}, 99 (1995); {\bf 94}, 343 (1995); 
K. Kato, S. Aoyama, S. Mukai, and K. Ikeda, Nucl. Phys. {\bf
A588}, c29 (1995).
\bibitem{unpub} A. Cs\'ot\'o, unpublished.
\bibitem{Janecke} J. J\"anecke {\it et al.}, Phys. Rev. C {\bf 54},
1070 (1996). 
\bibitem{Newton} R.~G. Newton, {\it Scattering theory of waves
and particles}, (Springer-Verlag, New York, 1982).
\bibitem{Cobisthesis} A. Cobis, PhD Thesis, University of
Aarhus, 1997.
\bibitem{Afnan} I.~R. Afnan, Aust. J. Phys. {\bf 44}, 201
(1991). 
\bibitem{3br} A. Cs\'ot\'o, Phys. Rev. C {\bf 49}, 2244 (1994). 
\bibitem{Glockle} W. Gl\"ockle, Phys. Rev. C {\bf 18}, 564
(1978). 
\bibitem{3n} A. Cs\'ot\'o, H. Oberhummer, and R. Pichler, Phys.
Rev. C {\bf 53}, 1589 (1996).
\end{references}
\end{document}